\documentclass[aps,prb,groupedaddress,amsmath,eqsecnum,twocolumn]{revtex4}
    \usepackage{bm}
    \usepackage{graphicx}

\newcommand\beq{\begin{equation}}
\newcommand\eeq{\end{equation}}
\newcommand\beqa{\begin{eqnarray}}
\newcommand\eeqa{\end{eqnarray}}
\newcommand{\nn}{\nonumber\\}

\begin{document}

\title{Virial coefficients, thermodynamic properties, and fluid-fluid transition of nonadditive hard-sphere mixtures}

\author{Andr\'es Santos}
\email{andres@unex.es}
\homepage{http://www.unex.es/eweb/fisteor/andres/}
\author{Mariano L\'opez de Haro}
\email{malopez@servidor.unam.mx}
\homepage{http://xml.cie.unam.mx/xml/tc/ft/mlh/}
\thanks{on sabbatical leave from Centro de Investigaci\'on en
Energ\'{\i}a, Universidad Nacional Aut\'onoma de M\'exico
(U.N.A.M.), Temixco, Morelos 62580, M{e}xico}

\author{Santos B. Yuste}
\email{santos@unex.es}\homepage{http://www.unex.es/fisteor/santos/}
\affiliation{Departamento de F\'{\i}sica, Universidad de
Extremadura, E-06071 Badajoz, Spain}

\date{\today}

\begin{abstract}
Different theoretical approaches for the thermodynamic properties
and the equation of state for  multicomponent mixtures of
nonadditive hard spheres in $d$ dimensions are presented in a
unified way. These include {the theory by Hamad, our previous formulation,} the original MIX1
theory, a recently proposed modified MIX1 theory, as well as a nonlinear extension
of the MIX1 theory proposed in this paper. Explicit expressions for the compressibility
factor, Helmholtz free energy, and second, third, and fourth virial
coefficients are provided. A comparison is carried out with recent
Monte Carlo data for the virial coefficients of asymmetric mixtures
and with available simulation data for the compressibility factor{, the critical consolute point, and the liquid-liquid coexistence  curves}.
The merits and limitations of each theory are pointed out.
\end{abstract}

\maketitle

\section{Introduction}
Nonadditive hard spheres represent a versatile model to study
various real physical systems. These include alloys, aqueous
electrolyte solutions, molten salts, rare gas mixtures, and colloids.
In these systems homocoordination and heterocoordination may be
interpreted in terms of excluded volume effects due to nonadditivity
of the repulsive (hard-core) part of the intermolecular potential
and so, for instance, the occurrence of liquid-liquid demixing in
real systems may be linked to a binary hard-sphere mixture with
positive nonadditivity, while negative nonadditivity may be invoked
to explain chemical short-range order in amorphous and liquid binary
mixtures with preferred heterocoordination. On the theoretical side,
prototype models of nonadditive hard-sphere mixtures such as the
Widom--Rowlinson model\cite{WR70} or the Asakura--Oosawa
model\cite{AO54} have been very useful to gain insight into
interesting physical aspects such as fluid-fluid phase transitions
and the nature of depletion forces.

A few years ago, in a paper\cite{SHY05} where a rather thorough
review of the theoretical and simulation work on nonadditive
hard-sphere mixtures was provided, we introduced an equation of
state of multicomponent nonadditive hard-sphere mixtures in $d$
dimensions. Such an equation of state results from a natural
extension of the one we had earlier proposed for additive hard
spheres,\cite{SYH99} has an explicit (simple) density dependence,
and by construction leads to the exact second and third virial
coefficients. In the case of $d=3$, in the same paper we compared
the predictions for the compressibility factor corresponding to our
proposal with those of the proposal by
Hamad,\cite{H94,H96a,H96b,H00} which shares some characteristics
with ours, and available simulation results for various binary
mixtures.\cite{RP94,JJR94a,JJR94b,JJR95,H97} We also compared the
predictions of the fourth and fifth virial coefficients arising
from the above two theoretical proposals and the then available
Monte Carlo results.\cite{SFG98,VM03} The restriction in the
comparison only to Hamad's approach was justified then by the fact
that Hamad had already proved that his proposal was superior to
other theories, including the so-called MIX1 theory originally due
to Melnick and Sawford.\cite{MS75}

Recently, Pellicane \textit{et al.}\cite{PCGS07} have reported new
evaluations of the fourth virial coefficient of a binary nonadditive
hard-sphere mixture covering a wide range of size ratios and values
of the nonadditivity parameter. Also recently, Paricaud\cite{P08} has proposed
a new equation of state for nonadditive hard-sphere mixtures which
is based on and corrects one of the deficiencies of the MIX1 theory,
namely the fact that MIX1 does not lead to the correct second virial
coefficient. These two recent papers serve as a motivation for the
present contribution. On the one hand, we want to see to what extent
the conclusions drawn from the analysis carried out in Ref.\
\onlinecite{SHY05} are still valid in view of the new available
data. On the other hand, we will also introduce a (new) nonlinear
extension of the MIX1 theory. As an extra bonus, we will write all
the theoretical expressions in a unified language which will
hopefully make the comparison much easier.

The paper is organized as follows. In order to make it
self-contained, in the next section we provide the necessary
background for the later development. Section \ref{sec3} provides
the explicit expressions for the contact values of the radial
distribution functions, compressibility factors, Helmholtz free
energies, and second, third, and fourth virial coefficients as given
by the original MIX1 theory, Paricaud's modified MIX1 theory, Hamad's
theory, and our earlier proposal.
A nonlinear extension of the MIX1
theory is also introduced here.
In Sec.\ \ref{sec4} we compare the
numerical values of the composition-independent virial coefficients,  compressibility factors, {and liquid-liquid coexistence curves for a variety of cases with available
Monte Carlo data}. The paper is closed in Sec.\ \ref{sec5} with some
concluding remarks.

\section{General background}
\label{sec2}

We consider an $N$-component mixture of nonadditive hard spheres in
$d$ dimensions. Let $\sigma_{ij}$ denote the hard-core distance of
the interaction between a sphere of species $i$ and a sphere of
species $j$. If the diameter of a sphere of species $i$ is
$\sigma_{i}\equiv \sigma_{ii}$, then
$\sigma_{ij}=\frac{1}{2}(\sigma_i+\sigma_j)(1+\Delta_{ij})$, where
$\Delta_{ij}\geq -1$ is a symmetric matrix with zero diagonal
elements ($\Delta_{ii}=0$) that characterizes the degree of
nonadditivity of the interactions. In the case of a binary mixture
($N=2$), the only nonadditivity parameter is
$\Delta=\Delta_{12}=\Delta_{21}$.

The compressibility factor $Z\equiv
p/\rho k_{B}T$ of the nonadditive mixture, where $\rho$ is the total number density, $p$ is the
pressure, $T$ is the temperature, and $k_B$ is the Boltzmann
constant, is given by
\beqa
Z(\rho,\{x_k\},\{\sigma_{k\ell}\})&=&1+2^{d-1}v_{d}\rho\sum_{i,j=1}^{N}x_ix_j
\sigma_{ij}^d \nn
&&\times g_{ij}(\rho,\{x_k\},\{\sigma_{k\ell}\}),
\label{1.0}
\eeqa
where $v_{d}=(\pi/4)^{d/2}/\Gamma(1+d/2)$ is the volume of a
$d$-dimensional sphere of unit diameter, $x_i=\rho_i/\rho$ is the
mole fraction of species $i$, $\rho_i$ being the partial number density
of particles of species $i$,
 and
$g_{ij}(\rho,\{x_k\},\{\sigma_{k\ell}\})\equiv g_{ij}(\rho)$ stands
for the radial distribution functions at contact. Unfortunately, no
general expression is known for $g_{ij}(\rho)$, but it may formally
be expanded in a power series in density as
\beqa
g_{ij}(\rho)&=&1+v_{d}\rho \sum_{k=1}^{N}x_{k}c_{k;ij}+(v_{d}\rho)^2\sum_{k,\ell=1}^{N}x_{k}x_\ell c_{k\ell;ij}
\nn
&& +{O}(\rho^3),
\label{1M}
\eeqa
where the coefficients $c_{k;ij}$, $c_{k\ell;ij}$, \ldots are
independent of the mole fractions but in general depend in a non
trivial way on the set of diameters $\{\sigma_{ij}\}$. To our
knowledge, only the coefficients linear in $\rho$ (\emph{i.e.},
$c_{k;ij}$) are known analytically for $d\le 3$. This formal series
expansion in the number density, Eq.\ (\ref{1M}), when substituted
into Eq.\ (\ref{1.0}), yields the virial expansion of $Z$ which we
write in the form
\begin{eqnarray}
Z(\rho)&=&1+\sum_{n=1}^\infty \rho^n B_{n+1} \nonumber\\
&=&1+\rho \sum_{i,j=1}^{N}x_{i}x_{j}B_{ij}+\rho^{2}\sum_{i,j,k=1}^{N}x_{i}x_{j}x_{k}B_{ijk}\nonumber\\
&&+\rho^{3}\sum_{i,j,k,\ell=1 }^{N}x_{i}x_{j}x_{k}x_{\ell}B_{ijk\ell
}+ {O}(\rho^4).
\label{2M}
\end{eqnarray}
Here $B_n$ is the usual $n$th virial coefficient of the multicomponent
mixture, which is a polynomial of degree $n$ in the mole fraction,  $B_{ij\cdots}$ being composition-independent coefficients. In terms of the coefficients $c_{k;ij}$ and $c_{k\ell;ij}$,
the composition-independent second, third, and fourth virial
coefficients are given by
\begin{equation}
B_{ij}=2^{d-1}v_{d}\sigma_{ij}^d,
\label{n1}
\end{equation}
\begin{equation}
B_{ijk}=\frac{2^{d-1}v_{d}^2}{3}\left(\sigma_{ij}^dc_{k;ij}+
\sigma_{ik}^dc_{j;ik}+\sigma_{jk}^dc_{i;jk}\right),
\label{n2}
\end{equation}
\begin{eqnarray}
B_{ijk\ell}&=&\frac{2^{d-1}v_{d}^3}{6}\left(\sigma_{ij}^dc_{k\ell;ij}+
\sigma_{ik}^dc_{j\ell;ik}+\sigma_{jk}^dc_{i\ell;jk}\right. \nonumber\\
&& \left.+ \sigma_{i\ell}^dc_{jk,i\ell}+\sigma_{j\ell}^dc_{ik,j\ell}
+\sigma_{k\ell}^dc_{ij;k\ell}\right).
\label{n3}
\end{eqnarray}

Along the path we have taken, the different theories for mixtures of
nonadditive hard spheres in $d$ dimensions may be related to
different proposals for $g_{ij}(\rho)$. In the next section we
provide the explicit expressions for the approximate proposals that
we will consider in this paper, including a new nonlinear extension of
the MIX1 theory.

\section{Some approximate theoretical developments\label{sec3}}

\subsection{MIX1 approximation}

The original MIX1 approximation,\cite{MS75} which we will indicate
with a superscript M, is equivalent to
\beqa
\sigma_{ij}^dg_{ij}^{\text{M}}(\rho)&=&\left(\frac{\sigma_i+\sigma_j}{2}\right)^d\left\{g^{\text{add}}_{ij}(\rho)
+Y_{ij}^\text{M}\frac{\partial}{\partial\rho}\left[\rho
g^{\text{add}}_{ij}(\rho)\right]\right\},
\nn
&&
\label{1.1}
\eeqa
where  $g^{\text{add}}_{ij}(\rho)$ are the contact values of the
\emph{additive} mixture and
\beq
Y_{ij}^\text{M}\equiv d\Delta_{ij}.
\label{1.2}
\eeq
Inserting Eq.\ \eqref{1.1} into Eq.\ \eqref{1.0} one gets
\beqa
Z^{\text{M}}(\rho)&=&
Z^{\text{add}}(\rho)+b_2 v_d
\rho\sum_{i,j}^N x_ix_j \left(\frac{\sigma_i+\sigma_j}{2}\right)^d
\nn
&&\times Y_{ij}^\text{M}\frac{\partial }{\partial\rho}\left[\rho
 g^{\text{add}}_{ij}(\rho)\right],
\label{1.3}
\eeqa
with $Z^{\text{add}}(\rho)$ the
compressibility factor of the \emph{additive} mixture with the same sets of mole fractions $\{x_k\}$ and diameters $\{\sigma_k\}$. The Helmholtz
free energy per particle in the MIX1 theory is then
\beqa
\frac{a^{\text{M}}(\rho)}{k_{B}T}&=& -1+\sum_{i}x_{i}\ln \left( x_{i}\rho
\lambda_{i}^d\right)+\frac{a^{\text{add}}_\text{ex}(\rho)}{k_{B}T}+b_2v_d\rho \nn
&&\times
\sum_{i,j} x_ix_j\left(\frac{\sigma_i+\sigma_j}{2}\right)^d
Y_{ij}^\text{M}g^{\text{add}}_{ij}(\rho),
\label{1.4}
\eeqa
where $\lambda_{i}$ is the de Broglie wavelength of particles of
species $i$, $a^{\text{add}}_\text{ex}(\rho)$ is the excess Helmholtz free
energy per particle of the \emph{additive} mixture and, for
convenience, we have identified  $2^{d-1}$ with the reduced second
virial coefficient in the one-component $d$-dimensional hard-sphere
fluid $b_2$. The second, third, and fourth virial coefficients of
the mixture are in turn given by
\beq
B_{ij}^{\text{M}}=b_2
v_d\left(\frac{\sigma_i+\sigma_j}{2}\right)^d
\left(1+Y_{ij}^\text{M}\right),
\label{1.5}
\eeq
\beqa
B_{ijk}^{\text{M}}&=&\frac{b_2v_d^2}{3}\left[
\left(\frac{\sigma_i+\sigma_j}{2}\right)^d
c^{\text{add}}_{k;ij}\left(1+2Y_{ij}^\text{M}\right)\right.
\nn &&
+\left(\frac{\sigma_i+\sigma_k}{2}\right)^d
c^{\text{add}}_{j;ik}\left(1+2Y_{ik}^\text{M}\right)
\nn
&&\left.+
\left(\frac{\sigma_j+\sigma_k}{2}\right)^d
c^{\text{add}}_{i;jk}\left(1+2Y_{jk}^\text{M}\right)\right],
\label{1.6}
\eeqa
\beqa
B_{ijk\ell}^{\text{M}}&=&\frac{b_2v_d^3}{6}\left[\left(\frac{\sigma_i+\sigma_j}{2}\right)^dc^{\text{add}}_{k\ell;ij}\left(1+3Y_{ij}^\text{M}\right)\right.\nn
&& +\left(\frac{\sigma_i+\sigma_k}{2}\right)^dc^{\text{add}}_{j\ell;ik}\left(1+3Y_{ik}^\text{M}\right)\nn
&& +\left(\frac{\sigma_i+\sigma_\ell}{2}\right)^dc^{\text{add}}_{jk;i\ell}\left(1+3Y_{i\ell}^\text{M}\right)\nn
&& +\left(\frac{\sigma_j+\sigma_k}{2}\right)^dc^{\text{add}}_{i\ell;jk}\left(1+3Y_{jk}^\text{M}\right)\nn
&& +\left(\frac{\sigma_j+\sigma_\ell}{2}\right)^dc^{\text{add}}_{ik;j\ell}\left(1+3Y_{j\ell}^\text{M}\right)\nn
&&\left. \left(\frac{\sigma_k+\sigma_\ell}{2}\right)^dc^{\text{add}}_{ij;k\ell}\left(1+3Y_{k\ell}^\text{M}\right)\right]
.
\label{1.7}
\eeqa
In Eqs.\ (\ref{1.6}) and (\ref{1.7}), $c^{\text{add}}_{k;ij}$ and
$c^{\text{add}}_{k\ell;ij}$ correspond to the coefficients in the
expansion of $g^{\text{add}}_{ij}(\rho)$ in powers of the number
density. Note that the second virial coefficient of the mixture in
the MIX1 theory is not exact [{compare Eqs.\ \eqref{n1} and \eqref{1.5}}], except
to first order in $\Delta_{ij}$. This problem can be traced back to
the fact that, according to Eq.\ \eqref{1.1},
\beq
\lim_{\rho\to
0}g_{ij}^{\text{M}}(\rho)=\frac{1+Y_{ij}^\text{M}}{\left(1+\Delta_{ij}\right)^d}\neq
1.
\label{1.7M}
\eeq
This is remedied by Paricaud's modification,\cite{P08} that is
described in the following subsection.

\subsection{Paricaud's modified MIX1 theory (mMIX1)}

In the modification of the MIX1 theory introduced recently by
Paricaud,\cite{P08} which we will refer to as mMIX1 and ascribe a superscript mM, one keeps
Eq.\ \eqref{1.1}, and hence Eqs.\ \eqref{1.3}--\eqref{1.7}, except
that $Y_{ij}^\text{M}$ is replaced by
\beq
Y_{ij}^\text{mM}\equiv(1+\Delta_{ij})^d-1.
\label{1.8}
\eeq
 With this change $Y_{ij}^\text{M}\to Y_{ij}^\text{mM}$, Eq.\ \eqref{1.1} becomes
\beqa
\sigma_{ij}^dg_{ij}^{\text{mM}}(\rho)&=&\left(\frac{\sigma_i+\sigma_j}{2}\right)^d\left\{g^{\text{add}}_{ij}(\rho)-
\frac{\partial}{\partial\rho}\left[\rho
g^{\text{add}}_{ij}(\rho)\right]\right\}\nn
&&+
\sigma_{ij}^d\frac{\partial}{\partial\rho}\left[\rho
g^{\text{add}}_{ij}(\rho)\right],
\label{1.1m}
\eeqa
or, equivalently,
\beq
g_{ij}^\text{mM}(\rho)=g^{\text{add}}_{ij}(\rho)+\frac{Y_{ij}^\text{mM}}{1+Y_{ij}^\text{mM}}\rho\frac{\partial}{\partial\rho}g^{\text{add}}_{ij}(\rho).
\label{3.1}
\eeq

In this way, instead of Eq.\ \eqref{1.7M}, we have $\lim_{\rho\to 0}g_{ij}^{\text{mM}}(\rho)= 1$ and thus the
second virial coefficient becomes exact. Otherwise, the third and
higher virial coefficients are still approximate. In particular, the third and fourth virial coefficients are given by Eqs.\ \eqref{1.6} and \eqref{1.7}, respectively, with $Y_{ij}^\text{M}\to Y_{ij}^\text{mM}$.

\subsection{Hamad's proposal}

Hamad's approximation,\cite{H94,H96a,H96b,H00} denoted by a
superscript H, consists of proposing the following ansatz
\begin{equation}
g_{ij}^{\text{H}}(\rho)=\left.g^{\text{pure}}(y)\right|_{y=\eta
X_{ij}^\text{H}},
\label{2.1}
\end{equation}
where $g^{\text{pure}}(y)$ is the contact value of the radial
distribution function of the one-component $d$-dimensional
hard-sphere fluid at the packing fraction $y$, $\eta\equiv v_d\rho
\langle\sigma^d\rangle$ is the packing fraction of the mixture (with
${\langle\sigma^m\rangle}=\sum_{i=1}^{N}x_i\sigma_i^m$), and
$X_{ij}^\text{H}$ will be specified later. {}From Eq.\ (\ref{2.1}) it
follows that the virial expansion of $g_{ij}(\rho)$ is given by
\beq
g_{ij}^{\text{H}}(\rho)=1+\sum_{n=1}^\infty
\frac{b_{n+2}}{b_2}\left(v_d\rho\langle\sigma^d\rangle
X_{ij}^\text{H}\right)^n,
\label{2.2}
\eeq
where $b_n$ is the reduced $n$th virial coefficient of the
one-component $d$-dimensional hard-sphere fluid. In particular,
comparing Eq.\ \eqref{2.2} with Eq.\ (\ref{1M}), one gets
\beq
\sum_kx_kc^\text{H}_{k;ij}=\frac{b_3}{b_2}\langle\sigma^d\rangle
X_{ij}^\text{H},
\label{2.3}
\eeq
\beq
\sum_{k,\ell}x_k x_\ell
c^\text{H}_{k\ell;ij}=\frac{b_4b_2}{b_3^2}\left(\sum_kx_kc^\text{H}_{k;ij}\right)^2,
\label{2.4a}
\eeq
so that
\beq
c^\text{H}_{k\ell;ij}=\frac{b_4b_2}{b_3^2}c^\text{H}_{k;ij}c^\text{H}_{\ell;ij}.
\label{2.4b}
\eeq
By requiring Eq.\ \eqref{2.1} to be exact to first order in density
(third virial coefficient), \emph{i.e.},
$c^\text{H}_{k;ij}=c_{k;ij}$, one must have
\beq
X_{ij}^\text{H}=\frac{b_2}{b_3}\frac{\sum_{k}x_kc_{k;ij}}{\langle
\sigma^d\rangle}.
\label{2.5}
\eeq

Using the above results, the compressibility factor and Helmholtz
free energy per particle in Hamad's proposal are given,
respectively, by
\beq
Z^{\text{H}}(\rho)=1+\sum_{i,j}\frac{x_i
x_j\sigma_{ij}^d}{\langle\sigma^d\rangle}
\frac{Z^{\text{pure}}\left(\eta
X_{ij}^\text{H}\right)-1}{X_{ij}^\text{H}}.
\label{2.7}
\eeq
and
\beqa
\frac{a^{\text{H}}(\rho)}{k_{B}T}&=& -1+\sum_{i}x_{i}\ln \left(
x_i\rho\lambda_{i}^d\right)\nn
&&+\sum_{i,j}\frac{x_ix_j\sigma_{ij}^d}{\langle\sigma^d\rangle
X_{ij}^\text{H}}\frac{a_{\text{ex}}^{\text{pure}}\left(\eta
X_{ij}^\text{H}\right)}{k_BT},
\label{2.8}
\eeqa
where $Z^{\text{pure}}(y)$ and $a_{\text{ex}}^{\text{pure}}(y)$ are
the compressibility factor and the excess Helmholtz free energy  per
particle, respectively, of the one-component $d$-dimensional hard-sphere fluid at
the packing fraction $y$. {}From Eqs.\ \eqref{n3} and
\eqref{2.4b} it follows that the fourth virial coefficient in
Hamad's approximation is
\beqa
{B}_{ijk\ell}^{\text{H}}&=&\frac{b_4b_2^2}{6b_3^2}v_d^3\left(\sigma_{ij}^d
c_{k;ij}c_{\ell;ij}+\sigma_{ik}^d
c_{j;ik}c_{\ell;ik}+\sigma_{i\ell}^d
c_{j;i\ell}c_{k;i\ell}\right.\nn &&\left.+\sigma_{jk}^d
c_{i;jk}c_{\ell;jk}+\sigma_{j\ell}^d
c_{i;j\ell}c_{k;j\ell}+\sigma_{k\ell}^d
c_{i;k\ell}c_{j;k\ell}\right).\nn
\label{2.6}
\eeqa
More in general, Eq.\ \eqref{2.7} yields
\beq
\label{virH}
B_n^\text{H}=b_n v_{d}^{n-1}\left(\frac{b_2}{b_3}\right)^{n-2}\sum_{i,j}x_i
x_j\sigma_{ij}^d \left(\sum_kx_kc_{k;ij}\right)^{n-2}.
\eeq

\subsection{The SHY proposal}

In Ref.\ \onlinecite{SHY05} we proposed the following ansatz for the
contact values of the radial distribution functions,
\begin{equation}
g_{ij}^{\text{SHY}}(\rho)= \frac{1}{1-\eta}
+\left[g^{\text{pure}}(\eta) -\frac{1}{1-\eta} \right] z_{ij},
\label{C1bis}
\end{equation}
where

\beq
z_{ij}=\left(\frac{b_3}{b_2}-1\right)^{-1}\left(\frac{\sum_k
x_kc_{k;ij}}{\langle\sigma^d\rangle}-1\right). \label{new1bis}
\eeq

This choice guarantees that $g_{ij}^{\text{SHY}}(\rho)$ is exact to
first order in density and thus this approximation retains the exact
second and third virial coefficients. When Eqs.\ (\ref{C1bis}) and
(\ref{new1bis}) are inserted into Eq.\ (\ref{1.0}) one gets
\begin{eqnarray}
Z^{\text{SHY}}(\rho)&=&1+\frac{b_3
B_2^* -b_2 B_3^*}{b_3-b_2}\frac{\eta}{1-\eta} \nonumber\\
&&+
\frac{B_3^*-
B_2^* }{b_3-b_2}\left[Z^{\text{pure}}(\eta)-1\right],
\label{new2}
\end{eqnarray}
where we have called $B_n^*\equiv B_n/(v_d\langle\sigma^d\rangle)^{n-1}$; note that $B_n^*\to b_n$ in the one-component limit.
In Eq.\ \eqref{new2} we have expressed
$Z^{\text{SHY}}(\rho)-1$ as a linear
combination of $\eta/(1-\eta)$ and $Z^{\text{pure}}(\eta)-1$, with
coefficients such that the second and third virial coefficients of
the mixture are exactly reproduced.
{}From the approximation (\ref{new2}), one may easily derive the
Helmholtz free energy per particle, which turns out to be
\begin{eqnarray}
\frac{a^{\text{SHY}}(\rho)}{ k_{B}T}&=&  -1+\sum_{i}x_{i}\ln \left(
x_i\rho\lambda_{i}^d\right)-\frac{b_3 B_2^*-b_2
B_3^*}{b_3-b_2}\nonumber\\
&& \times \ln(1-\eta)+
\frac{B_3^*- B_2^*}{b_3-b_2}\frac{a_{\text{ex}}^{\text{pure}}(\eta)}{
k_BT}.
\label{FEN-SYH}
\end{eqnarray}
 Also, Eq.\ (\ref{new2}) implies
that the $n$th virial coefficient is given by
\beq
{B_n^{\text{SHY}}}={\left(v_{d}\langle\sigma^d\rangle\right)^{n-1}}\left(\frac{b_n-b_2}{b_3-b_2}{B_3^*}-\frac{b_n-b_3}{b_3-b_2} {B_2^*}\right),
 \label{5.1}
\eeq
while for the composition-independent fourth virial coefficients one
gets the following explicit expression,
\begin{eqnarray}
B_{ijk\ell}^{\text{SHY}}&=&\frac{v_{d}(b_4-b_2)}{4(b_3-b_2)}\left(\sigma_i^d
B_{jk\ell}+\sigma_j^d B_{ik\ell}+\sigma_k^d
B_{ij\ell}\right.\nn
&&\left.+\sigma_\ell^d B_{ijk}\right)
-\frac{v_{d}^2(b_4-b_3)}{6(b_3-b_2)}\left(\sigma_i^d \sigma_j^d
B_{k\ell}\right.\nonumber\\
&&+\sigma_i^d \sigma_k^d B_{j\ell}+\sigma_i^d
\sigma_\ell^d B_{jk}+\sigma_j^d \sigma_k^d B_{i\ell}\nn
&&\left.+\sigma_j^d \sigma_\ell^d
B_{ik}+\sigma_k^d \sigma_\ell^d B_{ij}\right).
\label{5.2}
\end{eqnarray}

\subsection{A nonlinear MIX1 theory}

As a final theoretical proposal, in this subsection we introduce a
 new  extension of the MIX1 theory.

The SHY approximation, Eq.\ \eqref{C1bis},  is a ``local'' approximation with respect to density in the sense that the nonadditive contact value is expressed in terms of a reference contact value (here that of the one-component system) evaluated at precisely the \emph{same} density.
Analogously, both the original MIX1 approximation, Eq.\ \eqref{1.1}, and Paricaud's modified version, Eq.\ \eqref{3.1}, can be termed ``linearly non-local'' since the nonadditive contact value is furthermore expressed in terms of the \emph{first} density derivative of the additive contact value. In contrast, Hamad's approximation, Eq.\ \eqref{2.1}, is {``nonlinearly non-local''} because the reference contact value (again that of the one-component system) is taken at a totally \emph{different} scaled density.

 Our \emph{nonlinear} MIX1 (nlMIX1) approximation, labeled with nlM, is
inspired in both Eq.\ \eqref{3.1} and Eq.\ \eqref{2.1}. It consists
of assuming that
\beq
g_{ij}^\text{nlM}(\rho)=g^{\text{add}}_{ij}(\rho X_{ij}^\text{nlM}),
\label{3.2}
\eeq
where
\beq
X_{ij}^\text{nlM}\equiv
1+\frac{Y_{ij}^\text{mM}}{1+Y_{ij}^\text{mM}}.
\label{3.3}
\eeq
Expanding in powers of $X_{ij}^\text{nlM}-1$, Eq.\ \eqref{3.2} can
be formally rewritten as
\beqa
\label{gnl}
g_{ij}^\text{nlM}(\rho)&=&g^{\text{add}}_{ij}(\rho)+\sum_{n=1}^\infty
\frac{1}{n!}\left(\frac{Y_{ij}^\text{mM}}{1+Y_{ij}^\text{mM}}
\rho\right)^n \nn
&&\times\frac{\partial^n}{\partial
\rho^n}g^{\text{add}}_{ij}(\rho).
\eeqa
Comparison with Eq.\ \eqref{3.1} shows that
$g^{\text{mM}}_{ij}(\rho)$ can be seen as a first-order
approximation of $g^{\text{nlM}}_{ij}(\rho)$. Using Eq.\ \eqref{3.2}, the equation of state and Helmholtz
free energy per particle corresponding to the nlMIX1 theory are
given, respectively, by
\beq
Z^{\text{nlM}}(\rho)=1+b_2 v_d
\rho\sum_{i,j} x_i x_j \sigma_{ij}^d g^{\text{add}}_{ij}(\rho
X_{ij}^\text{nlM}),
\label{3.5}
\eeq
\beqa
\frac{a^{\text{nlM}}(\rho)}{k_{B}T}&=& -1+\sum_{i}x_{i}\ln \left(
x_i\rho\lambda_{i}^d\right)\nn
&&+b_2\sum_{i,j}\frac{x_ix_j\sigma_{ij}^d}{\langle\sigma^d\rangle X_{ij}^\text{nlM}} \mathcal{G}^{\text{add}}_{ij}(\rho
X_{ij}^\text{nlM}),\nn
\label{3.7}
\eeqa
where
\beq
\mathcal{G}^{\text{add}}_{ij}(\rho)\equiv v_d\langle\sigma^d\rangle \int_0^\rho d\rho' \,
g^{\text{add}}_{ij}(\rho').
\label{3.8}
\eeq

Note that, since $g_{ij}^{\text{mM}}(\rho)$ and $g_{ij}^{\text{nlM}}(\rho)$ coincide to
first order in density,  both give the same (approximate)
third virial coefficient, {namely Eq.\ \eqref{1.6} with $Y_{ij}^\text{M}\to Y_{ij}^\text{mM}$}. However, {the mMIX1 and nlMIX1 theories} differ at
the level of the fourth virial coefficient. In this case, instead of
Eq.\ \eqref{1.7} {[with $Y_{ij}^\text{M}\to Y_{ij}^\text{mM}$]} we have
\beqa
B_{ijk\ell}^\text{nlM}&=&\frac{b_2v_d^3}{6}\left[
\left(\frac{\sigma_i+\sigma_j}{2}\right)^dc^{\text{add}}_{k\ell;ij}
\frac{\left(1+2Y_{ij}^\text{mM}\right)^2}{1+Y_{ij}^\text{mM}}\right.\nn
&&+\left(\frac{\sigma_i+\sigma_k}{2}\right)^dc^{\text{add}}_{j\ell;ik}
\frac{\left(1+2Y_{ik}^\text{mM}\right)^2}{1+Y_{ik}^\text{mM}}\nn
&&+\left(\frac{\sigma_i+\sigma_\ell}{2}\right)^dc^{\text{add}}_{jk;i\ell}
\frac{\left(1+2Y_{i\ell}^\text{mM}\right)^2}{1+Y_{i\ell}^\text{mM}}\nn
&&+\left(\frac{\sigma_j+\sigma_k}{2}\right)^dc^{\text{add}}_{i\ell;jk}
\frac{\left(1+2Y_{jk}^\text{mM}\right)^2}{1+Y_{jk}^\text{mM}}\nn
&&+\left(\frac{\sigma_j+\sigma_\ell}{2}\right)^dc^{\text{add}}_{ik;j\ell}
\frac{\left(1+2Y_{j\ell}^\text{mM}\right)^2}{1+Y_{j\ell}^\text{mM}}\nn
&&\left.+\left(\frac{\sigma_k+\sigma_\ell}{2}\right)^dc^{\text{add}}_{ij;k\ell}
\frac{\left(1+2Y_{k\ell}^\text{mM}\right)^2}{1+Y_{k\ell}^\text{mM}}
\right].
\label{3.4}
\eeqa

It would be tempting to determine $X_{ij}^\text{nlM}$ in Eq.\ \eqref{3.2} by requiring agreement with the exact result to first order in density. This would give
\beq
X_{ij}\to\frac{\sum_k x_k c_{k;ij}}{\sum_k x_k c_{k;ij}^\text{add}}.
\label{3.9}
\eeq
Unfortunately, however, this implies a wrong composition dependence of the higher order terms in the expansion of $g_{ij}(\rho)$ in powers of $\rho$. In particular,
\beq
\sum_{k,\ell}x_kx_\ell c_{k\ell;ij}\to \left(\frac{\sum_k x_k c_{k;ij}}{\sum_k x_k c_{k;ij}^\text{add}}\right)^2
\sum_{k,\ell}x_kx_\ell c_{k\ell;ij}^\text{add}.
\label{3.10}
\eeq
While the left-hand side is quadratic in the mole fractions, the right-hand side is the ratio between a quartic function and a quadratic function.
In order to avoid inconsistencies as in \eqref{3.10} we need $X_{ij}^\text{nlM}$ to be independent of the mole fractions. Apart from that, $X_{ij}^\text{nlM}$ can be freely chosen but we will keep the choice \eqref{3.3} in order to make contact with the mMIX1 theory.

{Before closing this section, it is worth noting that, by construction, the nlMIX1 theory is \emph{a priori} not expected to be accurate for strong negative nonadditivities. This is because, on physical grounds, the parameter  $X_{ij}^\text{nlM}$ defined by Eq.\ \eqref{3.3} must be positive definite. This in turn implies, {from Eq.\ \eqref{1.8}}, the condition $\Delta_{ij}>-\left(1-2^{-1/d}\right)$. In the three-dimensional case, the above condition becomes $\Delta_{ij}\gtrsim -0.21$. As a matter of fact, the expansion \eqref{gnl} does not converge if $\Delta_{ij}\leq-\left(1-2^{-1/d}\right)$. Notwithstanding this,  from a practical point of view the nlMIX1 theory keeps providing meaningful results even if $\Delta_{ij}\leq-\left(1-2^{-1/d}\right)$, as will be seen in the next section.}

\section{Results}
\label{sec4}

Thus far the development has been rather general in the sense that
all the approximations we have discussed apply for any number of
components $N$ in the mixture and any dimensionality $d$. However,
it is only formal unless one specifies
$Z^{\text{add}}(\rho)$, $a^{\text{add}}_\text{ex}(\rho)$, and
$g^{\text{add}}_{ij}(\rho)$ in the case of all the MIX1 theories,
and $g^{\text{pure}}(y)$, $Z^{\text{pure}}(y)$, $a^{\text{pure}}_\text{ex}(y)$, and $c_{k;ij}$ in the cases of Hamad's and
the SHY approximations. In Ref.\ \onlinecite{SHY05} we introduced
for general $d$ the following approximation
\begin{equation}
c_{k;ij}=\sigma_{k;ij}^d+\left(\frac{b_3}{b_2}-1\right)\frac{\sigma_{i;jk}\sigma_{j;ik}}{\sigma_{ij}}
\sigma_{k;ij}^{d-1},
\label{n7}
\end{equation}
where
\begin{equation}
\sigma_{k;ij}\equiv\sigma_{ik}+\sigma_{jk}-\sigma_{ij}.
\label{diamekk}
\end{equation}
This is exact when $d=1$ and $d=3$ and proved to be accurate also
for $d=2$. We will also use it here.

As for the other remaining
quantities, since the new numerical data\cite{PCGS07} have been obtained for
$d=3$, we will restrict ourselves in the subsequent analysis only to
this dimensionality. Therefore in the MIX1 theories we will take for
$Z^{\text{add}}(\rho)$ and $a^{\text{add}}_\text{ex}(\rho)$ the expressions given by the popular
Boubl\'{\i}k--Mansoori--Carnahan--Starling--Leland (BMCSL) equation of
state,\cite{B70,MCSL71} namely
\beqa
Z^{\text{add}}(\rho)&=&\frac{1}{1-\eta
}+\frac{3\eta
}{(1-\eta
)^{2}}\frac{\langle\sigma\rangle\langle\sigma^2\rangle}{\langle\sigma^3\rangle}\nn
&&+\frac{\eta^{2}(3-\eta
)}{(1-\eta )^{3}}\frac{\langle\sigma^2\rangle^3}{\langle\sigma^3\rangle^2},
 \label{BMCSL}
\eeqa
\beqa
\frac{a_\text{ex}^\text{add}(\rho)}{k_{B}T}&=&-\ln(1-\eta)+\frac{3\eta
}{1-\eta}\frac{\langle\sigma\rangle\langle\sigma^2\rangle}{\langle\sigma^3\rangle}\nn
&&+\left[\frac{\eta}{(1-\eta)^2}+\ln(1-\eta)\right]\frac{\langle\sigma^2\rangle^3}{\langle\sigma^3\rangle^2},
\label{aBMCSL}
\eeqa
while for $g^{\text{add}}_{ij}(\rho)$ the choice will be the
Boubl\'{\i}k--Grundke--Henderson--Lee--Levesque (BGHLL)
values\cite{B70,GH72,LL73} given by
\beqa
g_{ij}^{\text{add}}(\rho)&=&\frac{1}{1-\eta }+\frac{3\eta }{(1-\eta
)^{2}}\frac{\sigma_i\sigma_j
\langle\sigma^2\rangle}{(\sigma_{i}+\sigma_{j})\langle\sigma^3\rangle}\nn
&&+\frac{2\eta^{2}}{(1-\eta)^{3}}\left[\frac{\sigma_i\sigma_j
\langle\sigma^2\rangle}{(\sigma_{i}+\sigma_{j})\langle\sigma^3\rangle}\right]^{2}.
\label{15BGHLL}
\eeqa
It follows from Eq.\ \eqref{15BGHLL} that $c^{\text{add}}_{k;ij}$
and $c^{\text{add}}_{k\ell;ij}$ are given by
\beq
\label{c3aad}
c^{\text{add}}_{k;ij}=\sigma_k^3+3\frac{\sigma_i\sigma_j}{\sigma_i+\sigma_j}\sigma_k^2,
\eeq
\beqa
\label{c4aad}
c_{k\ell;ij}^\text{add}&=&\sigma_k^3\sigma_\ell^3\left[1+3\frac{\sigma_i\sigma_j}{\sigma_i+\sigma_j}\frac{\sigma_k+\sigma_\ell}{\sigma_k\sigma_\ell}
+2\frac{\sigma_i^2\sigma_j^2}{\sigma_k\sigma_\ell(\sigma_i+\sigma_j)^2}\right].\nn
\eeqa
Equation \eqref{c3aad} is exact and agrees with Eq.\ \eqref{n7} in the three-dimensional additive limit ($b_3/b_2=\frac{5}{2}$, $\sigma_{k;ij}\to \sigma_k$). On the other hand, Eq.\ \eqref{c4aad} is approximate. According to Eq.\ \eqref{15BGHLL}, the quantity defined by Eq.\ \eqref{3.8} is given by
\beqa
\mathcal{G}_{ij}^{\text{add}}(\rho)&=&-\ln\left({1-\eta }\right)+3\left[\frac{\eta }{1-\eta
}+\ln\left({1-\eta }\right)\right]\nn
&&\times\frac{\sigma_i\sigma_j
\langle\sigma^2\rangle}{(\sigma_{i}+\sigma_{j})\langle\sigma^3\rangle}-2\Big[\frac{(1-3\eta/2)\eta}{(1-\eta)^{2}}\nn
&&+\ln\left({1-\eta }\right)\Big]\left[\frac{\sigma_i\sigma_j
\langle\sigma^2\rangle}{(\sigma_{i}+\sigma_{j})\langle\sigma^3\rangle}\right]^{2}.
\label{HijBGHLL}
\eeqa

Finally, in the case of the pure system, we will consider the
expressions corresponding to the Carnahan--Starling (CS) equation
of state,\cite{CS69} namely
\beq
g^{\text{pure}}(y)=\frac{1-y /2}{(1-y )^{3}},
\label{gCS}
\eeq
\beq
Z^{\text{pure}}(y)=\frac{1+y+y^2-y^3}{\left(1-y\right)^3}.
\label{CS}
\eeq
\beq
\frac{a^{\text{pure}}_\text{ex}(y)}{k_BT}=\frac{(4-3y)y}{\left(1-y\right)^2}.
\label{aCS}
\eeq

With the above choices, the five approximations reduce to the CS equation of state in the one-component case $\sigma_i=\sigma$. In the additive limit, however, there are three independent proposals: BMCSL (to which the original MIX1 theories and its two variants, mMIX1 and nlMIX1, reduce), Hamad's,  and what we referred to as eCS in Ref.\ \onlinecite{SYH99}. Of course, when nonadditivity is introduced, the five approximations differ from each other.

{\subsection{Virial coefficients}}

\begin{figure}
\includegraphics[width=0.87\columnwidth]{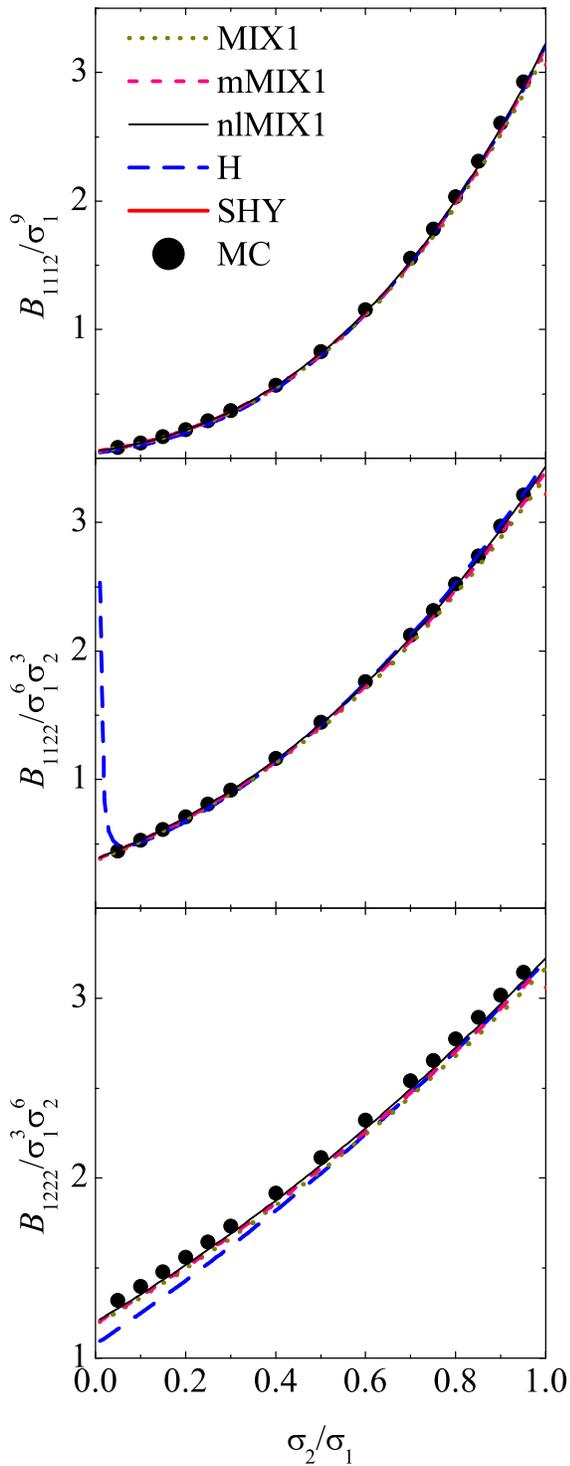}
\caption{Plot of the composition-independent fourth virial
coefficients $B_{1112}$, $B_{1122}$, and $B_{1222}$ versus the size
ratio $\sigma_2/\sigma_1$ for a nonadditivity parameter
$\Delta=0.05$. The dotted lines correspond to the original
MIX1 theory, Eq.\ \protect\eqref{1.7}, the  short-dash lines correspond to the mMIX1
theory, Eq.\ \protect\eqref{1.7} with $Y_{ij}^\text{M}\to Y_{ij}^\text{mM}$, the thin
solid lines correspond to the nlMIX1 theory, Eq.\ \protect\eqref{3.4}, the long-dash lines correspond to
Hamad's proposal,  Eq.\ \protect\eqref{2.6}, and the thick solid lines correspond to the SHY  proposal, Eq.\ \protect\eqref{5.2}. The symbols are Monte Carlo
data from Ref.\ \protect\onlinecite{PCGS07}.\label{fig1}}
\end{figure}
\begin{figure}
\includegraphics[width=0.87\columnwidth]{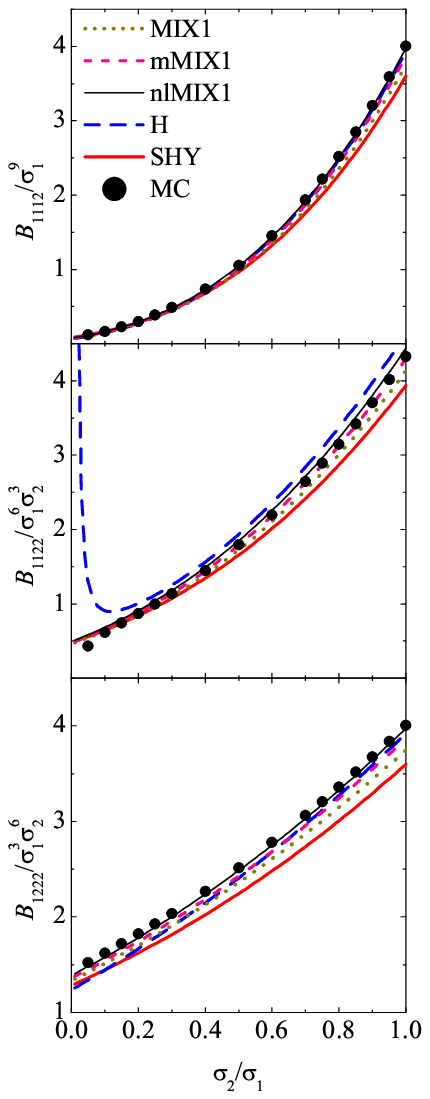}
\caption{Same as in  \protect\ref{fig1}, but for $\Delta=0.1$.
\label{fig2}}
\end{figure}
\begin{figure}
\includegraphics[width=0.87\columnwidth]{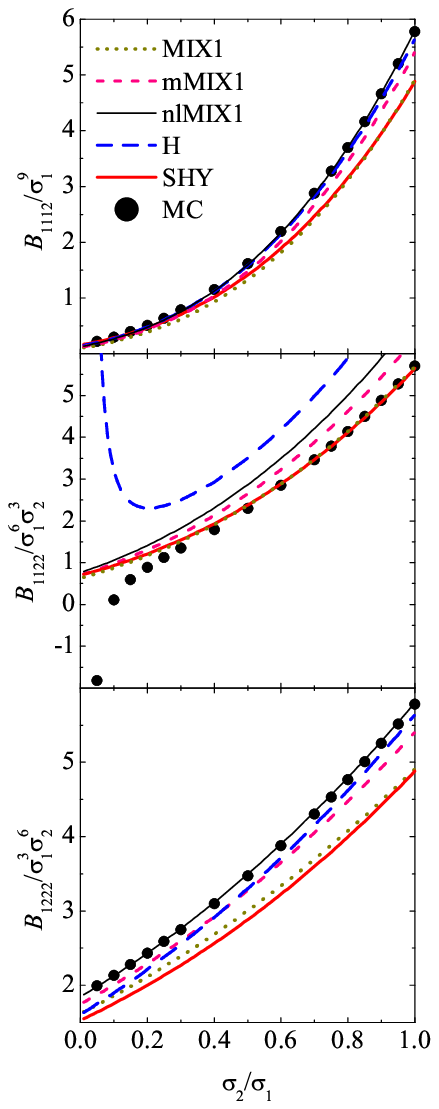}
\caption{Same as in  \protect\ref{fig1}, but for $\Delta=0.2$.
\label{fig3}}
\end{figure}
\begin{figure}
\includegraphics[width=0.87\columnwidth]{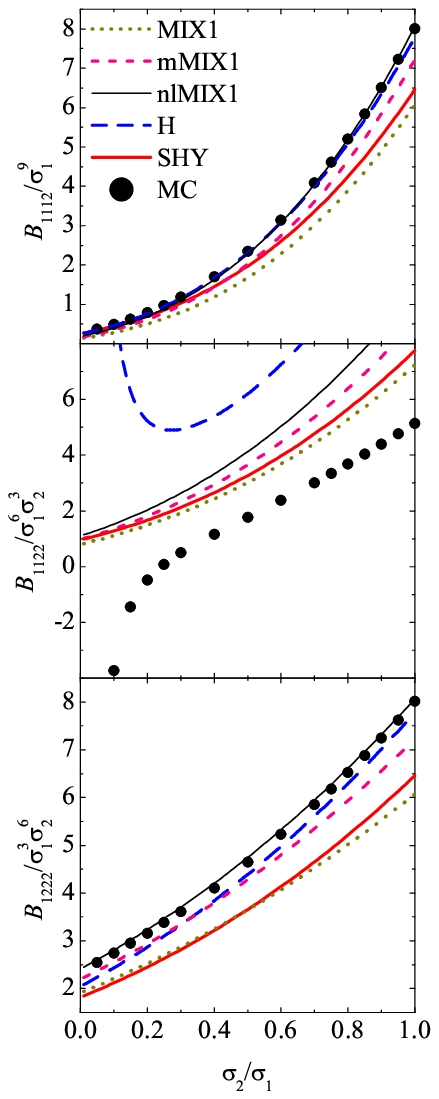}
\caption{Same as in  \protect\ref{fig1}, but for $\Delta=0.3$.
\label{fig4}}
\end{figure}
\begin{figure}
\includegraphics[width=0.87\columnwidth]{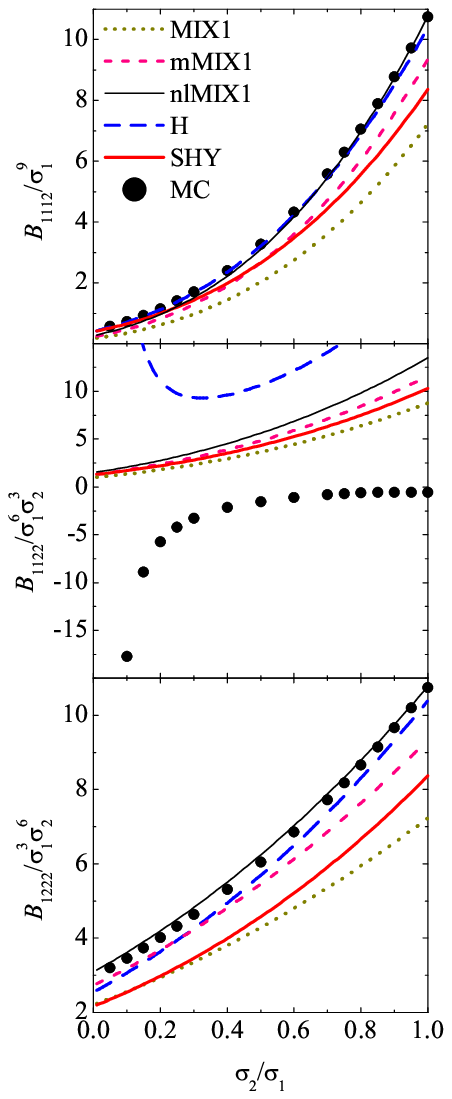}
\caption{Same as in \protect\ref{fig1}, but for $\Delta=0.4$.
\label{fig5}}
\end{figure}
\begin{figure}
\includegraphics[width=0.87\columnwidth]{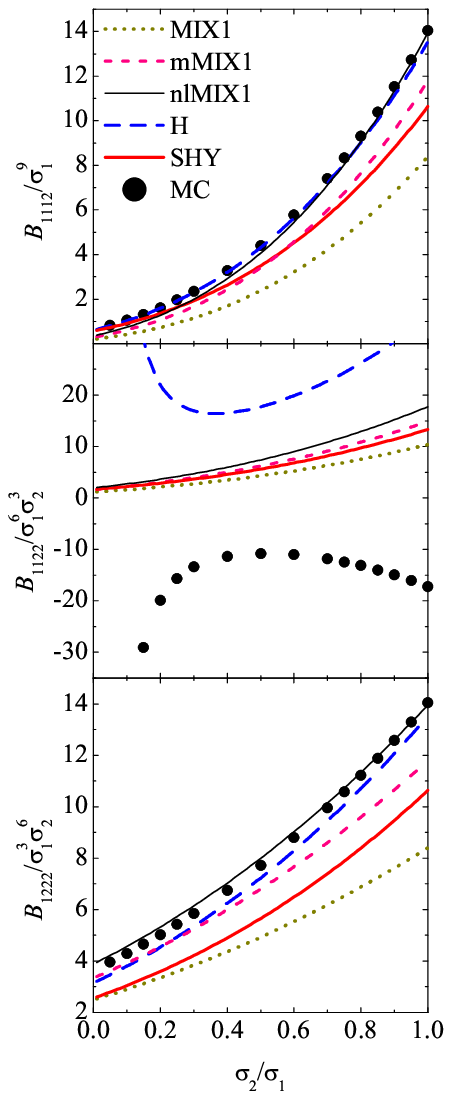}
\caption{Same as in \protect\ref{fig1}, but for $\Delta=0.5$.
\label{fig6}}
\end{figure}
\begin{figure}
\includegraphics[width=0.87\columnwidth]{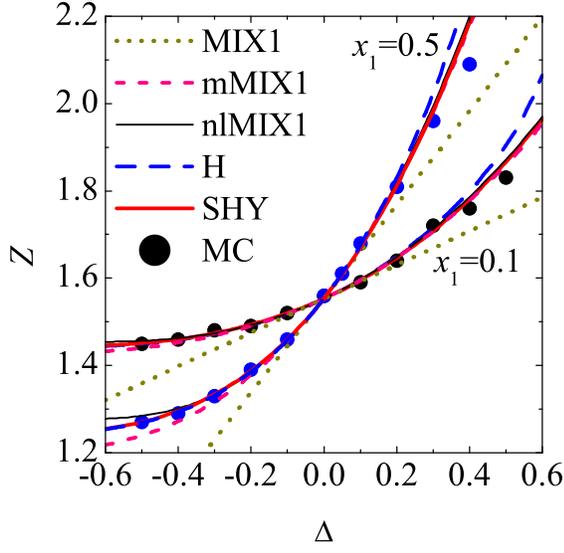}
\caption{Plot of the compressibility factor $Z$ versus the
nonadditivity parameter $\Delta$ for a symmetric binary mixture of
nonadditive hard spheres at $\eta=\pi/30$ and two different
compositions. The dotted lines correspond to the original
MIX1 theory, Eq.\ \protect\eqref{1.3}, the  short-dash lines correspond to the mMIX1
theory, Eq.\ \protect\eqref{1.3} with $Y_{ij}^\text{M}\to Y_{ij}^\text{mM}$, the thin
solid lines correspond to the nlMIX1 theory, Eq.\ \protect\eqref{3.5}, the long-dash lines correspond to
Hamad's proposal,  Eq.\ \protect\eqref{2.7}, and the thick solid lines correspond to the SHY  proposal, Eq.\ \protect\eqref{new2}.
The symbols are
results from Monte Carlo
simulations (Refs.\ \protect\onlinecite{JJR94a,JJR94b}).\label{fig7}}
\end{figure}
\begin{figure}
\includegraphics[width=0.87\columnwidth]{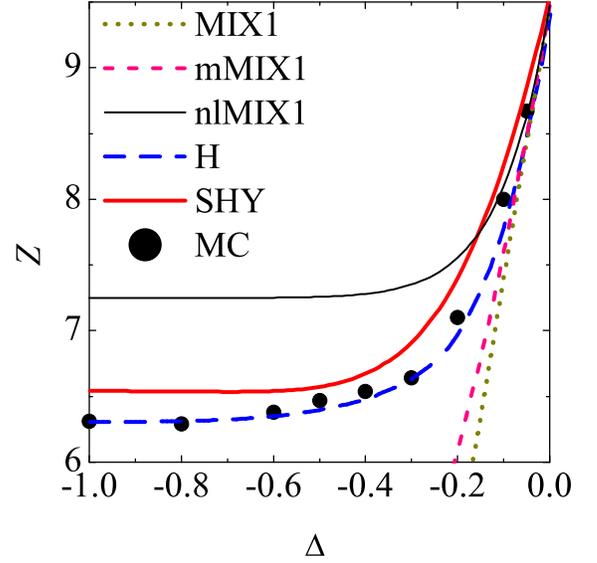}
\caption{Plot of the compressibility factor $Z$ versus the
nonadditivity parameter $\Delta$ for an equimolar asymmetric binary
mixture of nonadditive hard spheres with size ratio
$\sigma_{2}/\sigma_{1}=\frac{1}{3}$ at  $\eta=0.5$. The dotted line corresponds to the original
MIX1 theory, Eq.\ \protect\eqref{1.3}, the  short-dash line corresponds to the mMIX1
theory, Eq.\ \protect\eqref{1.3} with $Y_{ij}^\text{M}\to Y_{ij}^\text{mM}$, the thin
solid line corresponds to the nlMIX1 theory, Eq.\ \protect\eqref{3.5}, the long-dash line corresponds to
Hamad's proposal,  Eq.\ \protect\eqref{2.7}, and the thick solid line corresponds to the SHY  proposal, Eq.\ \protect\eqref{new2}
The symbols are
results from Monte Carlo
simulations (Ref.\ \protect\onlinecite{H97}).\label{fig8}}
\end{figure}
\begin{figure}
\includegraphics[width=0.87\columnwidth]{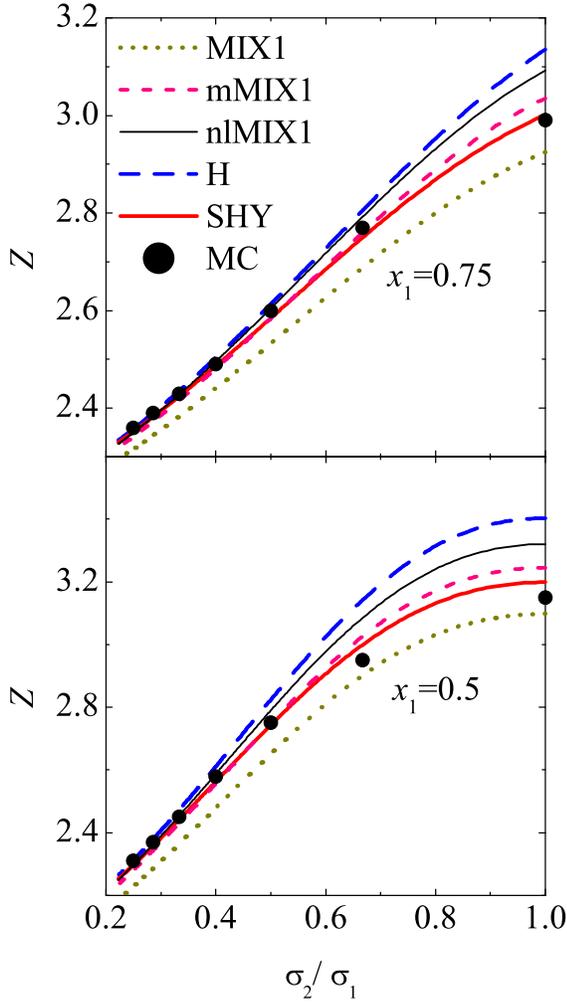}
\caption{Plot of the compressibility factor $Z$ versus the size ratio
$\sigma_{2}/\sigma_{1}$ for binary mixtures of nonadditive hard
spheres with $\Delta=0.2$ and $x_1=0.75$ (upper panel) and $x_1=0.5$
 (lower panel). The dotted lines correspond to the original
MIX1 theory, Eq.\ \protect\eqref{1.3}, the  short-dash lines correspond to the mMIX1
theory, Eq.\ \protect\eqref{1.3} with $Y_{ij}^\text{M}\to Y_{ij}^\text{mM}$, the thin
solid lines correspond to the nlMIX1 theory, Eq.\ \protect\eqref{3.5}, the long-dash lines correspond to
Hamad's proposal,  Eq.\ \protect\eqref{2.7}, and the thick solid lines correspond to the SHY  proposal, Eq.\ \protect\eqref{new2}
The symbols are
results from Monte Carlo
simulations (Ref.\ \protect\onlinecite{H97}).\label{fig9}}
\end{figure}
\begin{figure}
\includegraphics[width=0.87\columnwidth]{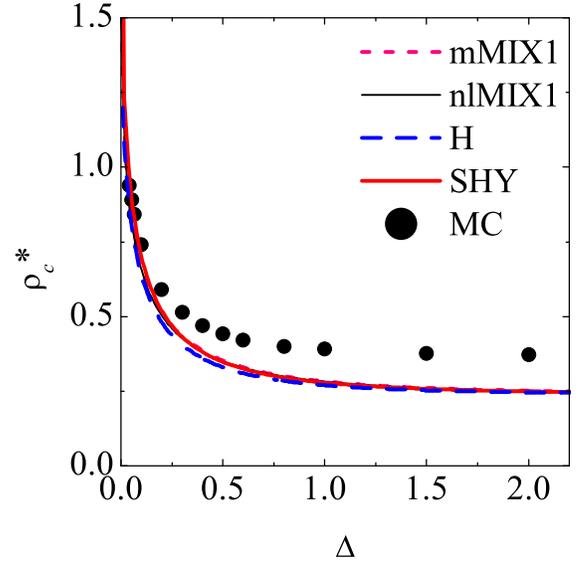}
\caption{Plot of the reduced critical density $\rho_c^*=\rho_c\sigma_\text{eff}^3$ versus the nonadditivity parameter
$\Delta$ for symmetric binary mixtures of nonadditive hard
spheres. The  short-dash line corresponds to the mMIX1
theory, the thin
solid line corresponds to the nlMIX1 theory, the long-dash line corresponds to
Hamad's proposal, and the thick solid line corresponds to the SHY  proposal.
The symbols are
results from Monte Carlo
simulations (Refs.\ \protect\onlinecite{JY03,G03,B05}).\label{fig10}}
\end{figure}
\begin{figure}
\includegraphics[width=0.87\columnwidth]{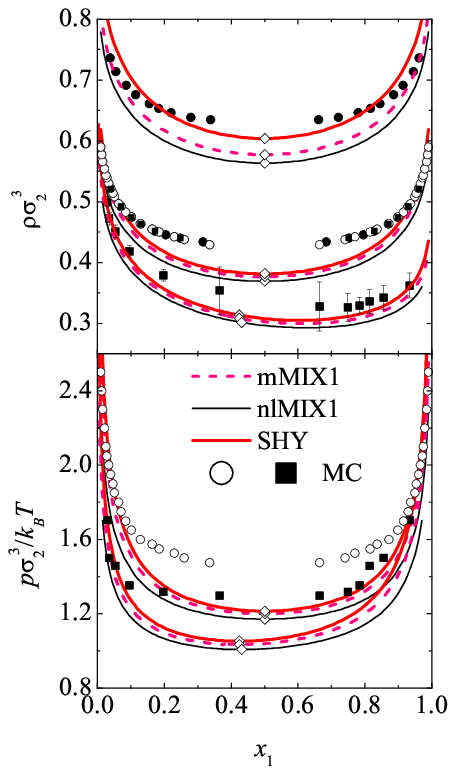}
\caption{{Liquid-liquid coexistence curves for several binary mixtures of nonadditive hard
spheres in the reduced density $\rho\sigma_2^3$ versus composition $x_1$ plane (top panel) and in the reduced pressure $p \sigma_2^3/k_B T$ versus composition $x_1$ plane (bottom panel).
{}From top to bottom, the set of curves correspond to  $(\sigma_2/\sigma_1,\Delta)=(1,0.1)$
 (absent in the bottom panel), $(1,0.2)$, and $(5/6,0.1818)$.  The  short-dash lines correspond to the mMIX1
theory, the thin
solid lines correspond to the nlMIX1 theory,  and the thick solid lines correspond to the SHY  proposal. The diamonds indicate the locations of the respective critical consolute points.
The other symbols are results from Monte Carlo
simulations: Ref.\ \protect\onlinecite{JY03} (filled circles), Ref.\ \protect\onlinecite{LALA96} (open circles), and Ref.\ \protect\onlinecite{RP94} (filled squares)}.\label{fig11}}
\end{figure}

Figures \ref{fig1}--\ref{fig6} show the comparison of the values
of the composition-independent fourth virial coefficients, as given
by the five theoretical proposals considered in this paper, with the
recent data of Pellicane  \emph{et al.}\cite{PCGS07,note}

One can immediately see that in the cases of $B_{1112}$ and
$B_{1222}$ the best overall performance is the one of the nlMIX1 theory, followed closely by Hamad's approximation.
Also worth
noting is that the mMIX1 theory already does a very good
job, especially for the smaller size ratios, while the original MIX1
theory gives the poorest agreement. As far as $B_{1122}$ is
concerned, the agreement of the theoretical predictions with the
Monte Carlo data is much less satisfactory, getting poorer as the
nonadditivity parameter is increased. Here, no approximation is
able to capture the negative values obtained by the Monte Carlo method for
$\Delta\ge 0.2$ and Hamad's approximation totally fails for small
size ratios, irrespective of the value of the nonadditivity
parameter. This is due to the fact that, while the four remaining theories correctly reproduce the scaling behavior
$B_{1122}\sim \sigma_1^6\sigma_2^3$ in the high-disparity limit $\sigma_2/\sigma_1\to 0$, Hamad's proposal yields $B_{1122}\sim \sigma_1^9$ in that limit. If one had to make a choice for this coefficient $B_{1122}$, either
the SHY proposal or the original MIX1 theory would perhaps be the
ones to go for (especially for $0\leq\Delta\leq 0.2$ and $0.3\leq \sigma_2/\sigma_1\leq 1$), but with all due reserves.

{One might reasonably wonder whether the use of more accurate expressions for the \emph{additive} contact values $g_{ij}^\text{add}$ might correct the inability of the theories examined in this paper to predict negative values of the virial coefficient $B_{1122}$ for small size ratio $\sigma_2/\sigma_1$ and large nonadditivity parameter $\Delta$. However, a closer analysis shows that this is not the case. According to Eq.\ \eqref{n3}, $B_{1122}\propto \sigma_1^3 c_{22;11}+\sigma_2^3 c_{11;22} +4\sigma_{12}^3 c_{12;12}$. Therefore, at least one of the \emph{nonadditive} second-order coefficients $c_{22;11}$, $c_{11;22}$, and $c_{12;12}$ must be negative if $B_{1122}<0$. In contrast, the additive coefficients $c_{ij;k\ell}^\text{add}$ are positive for any $\sigma_2/\sigma_1$ and, as a consequence, all the \emph{approximate} theories considered here predict positive values of $c_{ij;k\ell}$ for $\Delta>0$, as can be seen from Eqs.\ \eqref{1.1}, \eqref{3.1}, \eqref{2.1}, \eqref{C1bis}, and \eqref{3.2}.}

{\subsection{Compressibility factor}}

To complement the above information {on the virial coefficients}, in Figs.\ \ref{fig7}--\ref{fig9} we present the results of our calculations of the
compressibility factors of binary nonadditive hard-sphere mixtures
and a comparison with available simulation data.

Figure \ref{fig7} displays the dependence of $Z$ on the nonadditivity
parameter (both positive and negative) for a symmetric binary mixture at $\eta=\pi/30\simeq
0.105$ and two values of the mole fraction, namely $x_1=0.1$ and
$x_1=0.5$. In this case, both the SHY proposal and the nlMIX1
theory provide the best agreement, but the mMIX1 theory also
does a very good job. Hamad's proposal performs better at negative nonadditivities than at positive ones.
As for the MIX1 theory, being linear in $\Delta$, only captures the region of small $|\Delta|$.

The superiority of Hamad's theory for negative nonadditivities is confirmed by Fig.\ \ref{fig8}, which corresponds to
the case of an equimolar \emph{asymmetric} binary mixture with size ratio
$\sigma_2/\sigma_1=\frac{1}{3}$ and a packing fraction $\eta=0.5$. Here
Hamad's approximation clearly outperforms all the rest. As a matter of fact, it becomes exact in the extreme limit $\Delta\to -1$.\cite{SHY05}
A noteworthy feature
is that, in contrast with both the original MIX1 and the mMIX1
theories, the nlMIX1 theory at least captures correctly the
qualitative behavior of the compressibility factor with the  nonadditivity parameter for negative values and, in
particular, the initial decay. {This is remarkable in view of the fact that, as discussed at the end of the preceding section, the nlMIX1 theory is not expected to hold if $\Delta\lesssim -0.21$.}

Finally, in  Fig.\ \ref{fig9}
we present the results obtained for the size-ratio dependence of the
compressibility factor for $\eta=0.2$, a positive nonadditivity $\Delta=0.2$, and two compositions.
In agreement with the behavior observed in Fig.\ \ref{fig7} for $\Delta>0$, we see from  Fig.\ \ref{fig9} that   the
SHY is the superior theory also in the asymmetric case, although all the theories, with the exception of the MIX1, tend to coincide as the asymmetry increases. It is noteworthy that both the mMIX1 and the
nlMIX1 theories do a very reasonable job, better than Hamad's proposal.

{\subsection{Demixing}}

The availability of analytical expressions for the Helmholtz free energy per particle $a$ in all the previous theories {[cf.\ Eqs.\ \eqref{1.4}, \eqref{2.8}, \eqref{FEN-SYH}, and \eqref{3.7}]} may be exploited to address the problem of demixing in mixtures with positive nonadditivity. For simplicity, we will restrict ourselves here to binary mixtures. Since in these systems the temperature only plays the role of a scaling factor and a spinodal instability occurs, the mixture will phase separate into two liquid phases {(I and II) of different composition $x_1^\text{I}$ and $x_1^\text{II}$. For given size ratio $\sigma_2/\sigma_1$ and nonadditivity $\Delta$, by equating the pressure ($p^\text{I}=p^\text{II}$) and the two chemical potentials ($\mu_1^\text{I}=\mu_1^\text{II}$, $\mu_2^\text{I}=\mu_2^\text{II}$) of both phases, one may obtain  $\rho^\text{I}$, $\rho^\text{II}$, and $x_1^\text{II}$ as functions of $x_1^\text{I}$ and  thus derive the coexistence curve in the $\rho$-$x_1$ plane. The chemical potentials are defined by $\mu_i=\partial(\rho a)/\partial \rho_i$. In the binary case, this is equivalent to}
\beq
\mu_1=a+\frac{p}{\rho}+(1-x_1) a_x,\quad
\mu_2=a+\frac{p}{\rho}-x_1 a_x,
\label{4.1}
\eeq
{where $a_x\equiv (\partial a/\partial x_1)_\rho$. The two branches I and II of the coexistence line meet at the critical consolute point $(\rho_c,x_{1c})$, which can be determined by the two conditions}
\beq
0=\left(a_{\rho\rho}+\frac{2}{\rho}a_\rho\right)a_{xx}-a_{x\rho}^2,
\label{4.2}
\eeq
\beqa
0&=&a_{xxx}-3a_{xx\rho}\frac{a_{xx}}{a_{x\rho}}+3\left(a_{x\rho\rho}+\frac{2}{\rho}a_{x\rho}\right)\left(\frac{a_{xx}}{a_{x\rho}}\right)^2\nn
&&-
\left(a_{\rho\rho\rho}+\frac{6}{\rho}a_{\rho\rho}+\frac{6}{\rho^2}a_{\rho}\right)\left(\frac{a_{xx}}{a_{x\rho}}\right)^3.
\label{4.3}
\eeqa
{Here, as in Eq.\ \eqref{4.1}, each subscript $x$ or $\rho$ represents a derivative with respect to $x_1$ or $\rho$, respectively.}
{For symmetric mixtures, the critical composition is fixed, $x_{1c}=0.5$.}

{In Fig.\ \ref{fig10} we display the behavior of the {reduced critical}  density $\rho_c^*=\rho_c \sigma_\text{eff}^3$ in symmetric mixtures, {where $\sigma_\text{eff}^3\equiv \sum_{i,j}x_ix_j\sigma_{ij}^3$}, as a function of the nonadditivity parameter $\Delta$ for Hamad's theory, the SHY proposal, the mMIX1 and nlMIX1 theories, and the available simulation data. The original MIX1 theory has not been included since it has already been proved that it yields a poorer performance than Hamad's theory which is the least accurate in this instance. Note that all theoretical results underestimate $\rho_c^*$ and are very close to one another with perhaps a slightly better overall performance of the mMIX1 and the SHY.} {The use of the effective diameter $\sigma_\text{eff}$ to define the reduced critical density  in Fig.\ \ref{fig10} is motivated by the fact that $\rho_c^*$ is well defined for high nonadditivities, including the Widom--Rowlinson limit ($\sigma_1=\sigma_2\ll\sigma_{12}$ or $\Delta\to\infty$). }

{As far as the liquid-liquid coexistence curve is concerned, this may be represented in different thermodynamic planes. Here we have chosen the $\rho \sigma_2^3$-$x_1$ and the $p\sigma_2^3/ k_{B}T$-$x_1$ planes. Further, given the previous analysis concerning the comparison of the theoretical critical consolute points and simulation results, and the technical difficulties associated with the actual computation of the coexistence curves, only the results for the SHY, the mMIX1 and nlMIX1 theories will be presented. A comparison of available simulation results for liquid-liquid coexistence is done both for symmetric and asymmetric mixtures in Fig.\ \ref{fig11}, where the theoretical critical consolute points have also been included.
Notice that the qualitative trends observed in the simulations are well captured by all the theoretical developments, but in all instances they tend to underestimate the actual values of the reduced pressure and the reduced density along the coexistence. In particular, all theories correctly  predict that {the demixing transition occurs for lower densities as the nonadditivity parameter increases. Moreover, at a fixed value of $\Delta$ the coexistence densities (if measured in units of the diameter of the smaller spheres) decrease with increasing size asymmetry. Similar trends are observed for the pressure.} On the quantitative side, particularly in the density vs composition plane, albeit not very accurate, the SHY outperforms the other theoretical approximations.}

\section{Concluding remarks}
\label{sec5}

In this paper we have provided a self-contained presentation of
different theoretical developments to describe the thermodynamic
properties of nonadditive hard-core mixtures. In particular,
complementing the effort initiated in our previous paper on this
subject,\cite{SHY05} apart from repeating the SHY proposal and the
extension of Hamad's approach to general dimensionalities, here we
have provided extensions of the original MIX1 and Paricaud's modified MIX1 (mMIX1)
theories valid for all $d$.  We have introduced as well a new nonlinear
extension of the MIX1 (nlMIX1) theory, also valid for arbitrary $d$. In all
instances, explicit expressions have been provided for the contact
values of the radial distribution functions, the compressibility
factor, the Helmholtz free energy, and the second, third, and fourth
virial coefficients. The expressions  for $g_{ij}(\rho)$ and
$Z(\rho)$  are given in terms of either $g^{\text{add}}_{ij}(\rho)$
 and $Z^{\text{add}}(\rho)$ in the case of all the
MIX1 theories, or in terms of $g^{\text{pure}}(y)$, or equivalently
of  $Z^{\text{pure}}(y)= 1+2^{d-1}y
g^{\text{pure}}(y)$, in the cases of Hamad's and the SHY
approximations. For the sake of illustration and restricting to
three-dimensional systems ($d=3$), we have taken as input the BMCSL equation of state for
$Z^{\text{add}}(\rho)$ and the BGHLL contact values for
$g^{\text{add}}_{ij}(\rho)$ in the MIX1 theories, and the CS equation of
state for $Z^{\text{pure}}(y)$ in the SHY and {Hamad's} proposals.

To our knowledge, the idea of starting from the contact values of
the radial distribution functions in the case of the MIX1 theories
has not been considered before. This allowed us to construct the
nonlinear extension. Of course, while in the case of
mixtures the compressibility factor is determined uniquely once the
contact values of the radial distribution function are given, the
reciprocal is not true. Hence, the expressions we have provided for
these contact values are a further contribution of this work.

We have carried out {three} kinds of comparison between the five theories and ``exact'' numerical results.
First, the theoretical predictions of the composition-independent fourth virial coefficients have been tested against new available Monte Carlo data.\cite{PCGS07} In the cases of $B_{1112}$ and $B_{1222}$, the best overall agreement with the Monte Carlo values are obtained with the nlMIX1 theory, followed by Hamad's proposal. As for $B_{1122}$, none of the theories does well at high asymmetry and nonadditivity, the discrepancies being especially important in the case of Hamad's approximation.

As is well known, the first few virial coefficients are relevant to the equation of state in the low-density regime but not generally beyond it. Thus, in order to test the theoretical approaches at finite densities, we have made use of available simulation data for the compressibility factor.\cite{JJR94a,JJR94b,H97} The emerging scenario is that Hamad's approximation is excellent for negative nonadditivities, while the SHY proposal is the preferable one for positive nonadditivities.

Within the limited set {of compressibility factors}
that we have analyzed,  it is fair to say that the new nlMIX1
theory proposed in this paper is rather satisfactory and seems to be a good compromise
between accuracy and simplicity. Further assessment of this
assertion is precluded at this stage due to the scarcity of the
data. One of our hopes is therefore that the present paper may
encourage more work on the subject.

{Finally, the critical behavior and liquid-liquid coexistence of nonadditive
hard-sphere mixtures with positive nonadditivity has been examined. While the quantitative
agreement is not satisfactory, all theories seem
to capture correctly the qualitative trends obtained in the simulation. In this case our original SHY proposal gives the
best performance, but again the limited availability of data prevents us from carrying out a more thorough analysis. Once more we hope
that our findings may lead to the further needed work on this matter.}

\begin{acknowledgments}
This work  has been supported by the Ministerio de Educaci\'on y
Ciencia (Spain) through Grant No. FIS2007-60977 (partially financed
by FEDER funds).
\end{acknowledgments}


\begin{thebibliography}{99}
\bibitem{note} We note that  Tables 4 and 6 of Ref.\
\protect\onlinecite{PCGS07} contain some errata. The correct entries in the original
notation should be as follows (G. Pellicane and F. Saija, private comunication):
$D_{1122}=-2.0688(2)\times10^{-3}$ for $q=0.05$ and $\Delta=0.3$;
$D_{1122}=1.888(5) $ for $q=0.8$ and $\Delta=0.3$;
$D_{1122}=-5.6318(3)\times10^{-2}$ for $q=0.1$ and $\Delta=0.5$; and
$D_{1112}=9.314(9) $ for $q=0.8$ and $\Delta=0.5$.
\end{thebibliography}
\end{document}